\documentclass[final,3p,times]{elsarticle}

\usepackage{amsmath,amssymb}
\usepackage{graphicx}
\usepackage{booktabs}
\usepackage{tikz-cd}
\usetikzlibrary{arrows.meta,positioning}
\usepackage{enumitem}

\usepackage{xurl}
\usepackage[
  colorlinks=true,
  citecolor=blue,
  urlcolor=blue,
  linkcolor=black
]{hyperref}

\journal{Journal Name}

\begin{document}
\begin{frontmatter}

\title{Attributing Differences Between Forecast Runs to Input Changes: Applications to CCAR and CECL}

\author[aff1]{Xuan Mei\fnref{fn1}}
\ead{xuan.mei@chase.com}

\author[aff1]{Junze Lin\fnref{fn1}}
\ead{junze.lin@jpmorgan.com} 

\affiliation[aff1]{
  organization={JPMorgan Chase \& Co.},
  address={545 Washington Blvd.},
  city={Jersey City},
  state={NJ},
  postcode={07310},
  country={USA}
}
\fntext[fn1]{The authors work in the Wholesale Credit QR group at JPMorgan Chase \& Co. This article was prepared in their personal capacities. The views expressed are solely those of the authors and do not represent those of JPMorgan Chase \& Co.}


\begin{abstract}
Forecasting systems used in the Comprehensive Capital Analysis and Review
(CCAR) and Current Expected Credit Losses (CECL) processes combine portfolio
data, macroeconomic scenarios, model specifications, business assumptions,
and management adjustments. When the forecast changes from one run to the
next, practitioners need an attribution that reconciles to the total change
without depending on an arbitrary sequence of input replacements. This paper
formulates forecast-gap attribution as a cooperative game and examines several
approaches: the exact Shapley value, hierarchical or nested Shapley values,
Integrated Gradients, Gradient SHAP, Permutation SHAP, and Kernel SHAP. We
compare their allocation rules, computational costs, implementation
requirements, and limitations in production forecasting systems. The analysis
provides a practical framework for choosing an attribution method according to
the number and type of inputs, the feasibility of hybrid forecast runs, and
the need for interpretability, reproducibility, and governance.
\end{abstract}


\end{frontmatter}

\section{Introduction}
\label{sec:introduction}

U.S. banks rely on forward-looking forecasting frameworks for regulatory
capital planning and financial reporting. The Comprehensive Capital Analysis
and Review (CCAR) is the Federal Reserve's assessment of the capital adequacy
and capital-planning practices of large banking organizations
\cite{federalreserve-ccar-qa}. Under this process, projected revenues, credit
losses, expenses, and capital ratios are evaluated under baseline and severely
adverse macroeconomic scenarios to assess whether a bank could absorb losses
while continuing to lend. The Current Expected Credit Losses (CECL) framework,
by contrast, is a U.S. GAAP accounting standard that requires institutions to
recognize an allowance for expected credit losses over the contractual life of
financial assets measured at amortized cost. The estimate incorporates
historical experience, current conditions, and reasonable and supportable
forecasts \cite{occ-allowances-2026}.

Although CCAR and CECL differ in purpose, horizon, and governing requirements,
both typically rely on complex forecasting systems that combine multiple
models and data sources to estimate future credit losses. When two runs---for
example, CCAR 2025 and CCAR 2026, or the February and May 2026 CECL
runs---produce different forecasts, risk managers need to understand how much
of the difference is attributable to each change in the inputs.

The precise configuration varies across institutions, but the principal inputs
to these systems generally include:
\begin{itemize}
  \item a portfolio launch-point (LP) file containing account- or loan-level data, such as balances, borrower
        characteristics, collateral values, and credit ratings;
  \item hypothetical scenarios expressed through macroeconomic variables (MEVs), including paths for the unemployment rate (UER), gross domestic product (GDP), the housing price index (HPI), interest rates, and commercial-property net operating income (NOI) and price indices;
  \item the specifications, parameters, and production-code versions of models for probability of default (PD), loss given default (LGD), exposure at default (EAD), prepayment, and rating migration (RM);
  \item business assumptions, including planned originations, repayments,
        line utilization, and portfolio growth; and
  \item expert judgments, management overlays, and other post-model
        adjustments.
\end{itemize}

Each of these inputs may change between runs. Consider two consecutive annual
CCAR exercises. The earlier exercise may use a launch-point portfolio dated
December 31, 2024, whereas the later exercise uses a portfolio dated December
31, 2025. Over that year, the portfolio changes through prepayments, defaults,
maturities, and new originations; collateral values also move with market
conditions. The macroeconomic scenario reflects a new economic outlook,
models may be updated, and business assumptions or management overlays may be
revised. Because several changes occur at once, a defensible method is needed
to separate their effects on the forecast.

To formulate this problem mathematically, let $F(x_1, \ldots, x_k)$ denote
the forecasting system, where the $x_i$ are its individual inputs. Let
$(x^{(0)}_1, \ldots, x^{(0)}_k)$ and
$(x^{(1)}_1, \ldots, x^{(1)}_k)$ denote the inputs to the first and second
runs, respectively. The total forecast gap is defined as
\begin{equation}
  \Delta \triangleq
  F(x^{(1)}_1, \ldots, x^{(1)}_k)
  - F(x^{(0)}_1, \ldots, x^{(0)}_k).
  \label{eq:total-gap}
\end{equation}
The problem is to attribute $\Delta$ to the changes in the individual inputs.

Because the forecasting system $F$ is typically complex (see, for example,
Mei and Lin~\cite{mei2026attributing} for an expected-loss framework),
practitioners often use walk analysis. Starting with the first-run inputs, the
analyst replaces one input at a time with its second-run value and reruns the
system after each replacement. The contribution assigned to an input is the
change in the forecast at the step when that input is replaced. These
increments telescope, so their sum equals the total forecast gap $\Delta$.
The procedure can be formalized as follows.

\begin{figure}[htbp]
  \centering
  \resizebox{\textwidth}{!}{%
  \begin{tikzpicture}[
    >=Latex,
    state/.style={
      draw=blue!65!black,
      rounded corners=3pt,
      fill=blue!7,
      thick,
      align=center,
      minimum width=3.5cm,
      minimum height=1.8cm,
      inner sep=6pt
    },
    endpoint/.style={state,fill=orange!15,draw=orange!70!black},
    walkarrow/.style={->,very thick,draw=black!70},
    total/.style={->,thick,draw=orange!80!black}
  ]
    \node[endpoint] (s0) at (0,0) {
      \textbf{First run (baseline)}\\
      $\boldsymbol{x}^{\pi,0}=\boldsymbol{x}^{(0)}$\\[1mm]
      $F(\boldsymbol{x}^{\pi,0})$
    };

    \node[state] (s1) at (5.0,0) {
      \textbf{Hybrid run 1}\\
      $\boldsymbol{x}^{\pi,1}$\\[1mm]
      $F(\boldsymbol{x}^{\pi,1})$
    };

    \node[state] (sj) at (10.0,0) {
      \textbf{Hybrid run $j$}\\
      $\boldsymbol{x}^{\pi,j}$\\[1mm]
      $F(\boldsymbol{x}^{\pi,j})$
    };

    \node[endpoint] (sk) at (15.0,0) {
      \textbf{Second run}\\
      $\boldsymbol{x}^{\pi,k}=\boldsymbol{x}^{(1)}$\\[1mm]
      $F(\boldsymbol{x}^{\pi,k})$
    };

    \draw[walkarrow] (s0) --
      node[above,font=\small] {$C_{\pi_1}^{\pi}$} (s1);

    \draw[walkarrow] (s1) --
      node[above,align=center,font=\small]
      {$C_{\pi_2}^{\pi},\ldots,C_{\pi_j}^{\pi}$} (sj);

    \draw[walkarrow] (sj) --
      node[above,align=center,font=\small]
      {$C_{\pi_{j+1}}^{\pi},\ldots,C_{\pi_k}^{\pi}$} (sk);

    \draw[total] (s0.south) -- ++(0,-1.0) --
      node[below,align=center,font=\small]
      {$\displaystyle\sum_{r=1}^{k}C_{\pi_r}^{\pi}=\Delta$}
      ++(15.0,0) -- (sk.south);
  \end{tikzpicture}%
  }
  \caption{Walk analysis for an ordering $\pi$. Starting from the first-run
  inputs, one input is replaced at each underlying step. The resulting forecast
  changes telescope to the total gap between the two runs.}
  \label{fig:walk-analysis}
\end{figure}
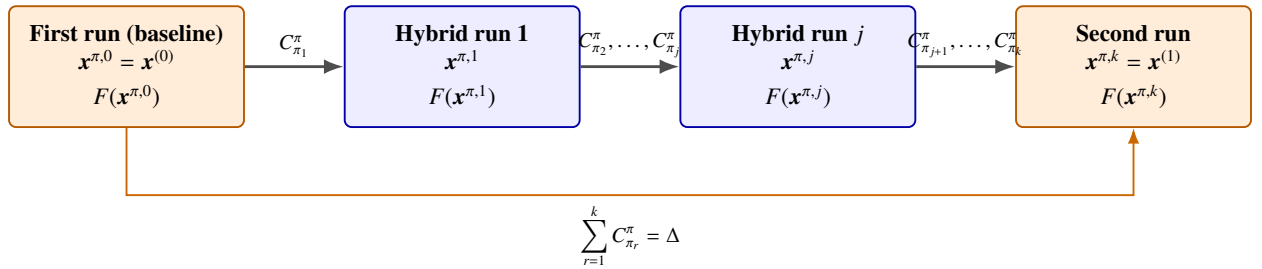

Let
$\pi=(\pi_1,\ldots,\pi_k)$ be a permutation of $\{1,\ldots,k\}$ specifying the
order in which the inputs are changed. For $j=0,\ldots,k$, define the hybrid
input vector $\boldsymbol{x}^{\pi,j}$ componentwise by
\begin{equation}
  x_i^{\pi,j}
  \triangleq
  \begin{cases}
    x_i^{(1)}, & i\in\{\pi_1,\ldots,\pi_j\},\\
    x_i^{(0)}, & i\notin\{\pi_1,\ldots,\pi_j\}.
  \end{cases}
  \label{eq:hybrid-input}
\end{equation}
Thus, $\boldsymbol{x}^{\pi,0}=\boldsymbol{x}^{(0)}$ and
$\boldsymbol{x}^{\pi,k}=\boldsymbol{x}^{(1)}$. The contribution assigned to
the input changed at step $j$ is
\begin{equation}
  C_{\pi_j}^{\pi}
  \triangleq
  F\!\left(\boldsymbol{x}^{\pi,j}\right)
  -F\!\left(\boldsymbol{x}^{\pi,j-1}\right),
  \qquad j=1,\ldots,k.
  \label{eq:walk-contribution}
\end{equation}
Figure~\ref{fig:walk-analysis} illustrates this sequential replacement process.
At each underlying step, exactly one input is changed and the resulting
forecast change is assigned to that input; intermediate hybrid runs are
suppressed in the diagram for compactness.
Summing the contributions yields the telescoping identity
\begin{align}
  \sum_{j=1}^{k} C_{\pi_j}^{\pi}
  &= \sum_{j=1}^{k}
     \left[
       F\!\left(\boldsymbol{x}^{\pi,j}\right)
       -F\!\left(\boldsymbol{x}^{\pi,j-1}\right)
     \right] \notag\\
  &=F\!\left(\boldsymbol{x}^{(1)}\right)
    -F\!\left(\boldsymbol{x}^{(0)}\right)
   =\Delta.
  \label{eq:walk-telescope}
\end{align}
Thus, every ordering $\pi$ gives an exact decomposition of the total gap.
However, CCAR and CECL systems are nonlinear and contain interactions among
their inputs, so $C_i^{\pi}$ generally varies with $\pi$. A conventional walk
is exact along the selected path but is not order invariant: changing the
sequence of replacements can materially change the amount attributed to each
input $x_i$.

The remainder of this paper examines attribution methods that remove this order or path dependence.

\section{Methods}
\subsection{Exact Shapley Value Approach}
\subsubsection{Methodology}
A single walk allocates interaction effects according to its chosen order. An
order-independent alternative is to consider all $k!$ permutations and average
the marginal contribution of each input across them. The resulting allocation
is the Shapley value, introduced by Shapley~\cite{shapley1953value} as a
principled allocation rule in cooperative game theory.

To express the problem using the traditional cooperative-game convention, let
$N=\{1,\ldots,k\}$ be the set of players, where player $i$ represents input
$x_i$. A coalition $S\subseteq N$ is interpreted as the set of players that
have already played. When player $i$ plays, the value of its corresponding
input is switched from $x_i^{(0)}$ to $x_i^{(1)}$. Thus, for each coalition
$S$, define the hybrid input vector $\boldsymbol{x}^{S}$ by
\begin{equation}
  x_i^{S}
  \triangleq
  \begin{cases}
    x_i^{(1)}, & i\in S,\\
    x_i^{(0)}, & i\notin S.
  \end{cases}
  \label{eq:coalition-input}
\end{equation}
The value of coalition $S$ is the change in the forecast produced by the
players in $S$ relative to the first-run baseline:
\begin{equation}
  v(S)
  \triangleq
  F\!\left(\boldsymbol{x}^{S}\right)
  -F\!\left(\boldsymbol{x}^{\varnothing}\right).
  \label{eq:coalition-value}
\end{equation}
Consequently, $v(\varnothing)=0$ and $v(N)=\Delta$.

Let $\mathcal{S}_k$ denote the set of all permutations of the players. For a
permutation $\pi\in\mathcal{S}_k$, define the predecessor coalition of player
$i$ as
\begin{equation}
  P_i^{\pi}
  \triangleq
  \left\{j\in N:\pi^{-1}(j)<\pi^{-1}(i)\right\},
  \label{eq:predecessor-coalition}
\end{equation}
which contains all players that appear before player $i$ in the ordering
$\pi$. The marginal contribution of player $i$ when it joins this coalition is
\begin{equation}
  m_i^{\pi}
  \triangleq
  v\!\left(P_i^{\pi}\cup\{i\}\right)
  -v\!\left(P_i^{\pi}\right).
  \label{eq:player-marginal-contribution}
\end{equation}
This quantity is identical to the walk contribution $C_i^{\pi}$ in
Equation~\eqref{eq:walk-contribution}. For player $i$ (equivalently, input
$i$), the Shapley attribution is the average marginal contribution over all
possible orders:
\begin{equation}
  \phi_i
  \triangleq
  \frac{1}{k!}
  \sum_{\pi\in\mathcal{S}_k} m_i^{\pi}
  =\frac{1}{k!}
  \sum_{\pi\in\mathcal{S}_k} C_i^{\pi},
  \qquad i=1,\ldots,k.
  \label{eq:shapley-permutation}
\end{equation}
Equivalently, the Shapley value can be written in its more familiar
coalition-set form:
\begin{equation}
  \phi_i
  =
  \sum_{S\subseteq N\setminus\{i\}}
  \frac{|S|!\,(k-|S|-1)!}{k!}
  \left[
    v\!\left(S\cup\{i\}\right)-v(S)
  \right].
  \label{eq:shapley-coalition}
\end{equation}
For a given coalition $S$, there are $|S|!$ ways to order its members before
player $i$ and $(k-|S|-1)!$ ways to order the remaining players after player
$i$. Hence, the coefficient in Equation~\eqref{eq:shapley-coalition} is the
fraction of all $k!$ permutations for which $S$ is exactly the predecessor
coalition of player $i$, establishing the equivalence between the set and
permutation formulations.

\begin{figure*}[htbp]
  \centering
  \begin{tikzpicture}[
    >=Latex,
    coalition/.style={
      draw=blue!60!black,
      rounded corners=2pt,
      fill=blue!6,
      thick,
      align=center,
      minimum width=2.15cm,
      minimum height=0.85cm,
      inner sep=3pt
    },
    endpoint/.style={coalition,fill=orange!14,draw=orange!75!black},
    ordinary/.style={->,draw=black!42,semithick},
    player/.style={->,draw=red!75!black,very thick},
    weight/.style={font=\small,align=center,text=red!70!black}
  ]
    \node[endpoint] (empty) at (0,0) {$\varnothing$\\$v(\varnothing)=0$};

    \node[coalition] (one)   at (4.0, 2.0) {$\{1\}$\\$v(\{1\})$};
    \node[coalition] (two)   at (4.0, 0.0) {$\{2\}$\\$v(\{2\})$};
    \node[coalition] (three) at (4.0,-2.0) {$\{3\}$\\$v(\{3\})$};

    \node[coalition] (onetwo)   at (8.0, 2.0) {$\{1,2\}$\\$v(\{1,2\})$};
    \node[coalition] (onethree) at (8.0, 0.0) {$\{1,3\}$\\$v(\{1,3\})$};
    \node[coalition] (twothree) at (8.0,-2.0) {$\{2,3\}$\\$v(\{2,3\})$};

    \node[endpoint] (all) at (12.0,0) {$\{1,2,3\}$\\$v(N)=\Delta$};

    \draw[ordinary] (empty) -- (one);
    \draw[ordinary] (empty) -- (three);
    \draw[ordinary] (one) -- (onethree);
    \draw[ordinary] (three) -- (onethree);
    \draw[ordinary] (two) -- (onetwo);
    \draw[ordinary] (two) -- (twothree);
    \draw[ordinary] (onetwo) -- (all);
    \draw[ordinary] (twothree) -- (all);

    \draw[player] (empty) -- node[above,font=\small]
      {$v(\{2\})-v(\varnothing)$} (two);
    \draw[player] (one) -- node[above,font=\small]
      {$v(\{1,2\})-v(\{1\})$} (onetwo);
    \draw[player] (three) -- node[below,font=\small]
      {$v(\{2,3\})-v(\{3\})$} (twothree);
    \draw[player] (onethree) -- node[above,font=\small]
      {$v(N)-v(\{1,3\})$} (all);

    \node[weight] at (6.0,-3.15) {For input 2:\quad
      $\displaystyle
      \phi_2=
      \frac{1}{3}\bigl[v(\{2\})-v(\varnothing)\bigr]
      +\frac{1}{6}\bigl[v(\{1,2\})-v(\{1\})\bigr]$\\[1mm]
      $\displaystyle\hphantom{\phi_2={}}
      +\frac{1}{6}\bigl[v(\{2,3\})-v(\{3\})\bigr]
      +\frac{1}{3}\bigl[v(N)-v(\{1,3\})\bigr].$};
  \end{tikzpicture}
  \caption{Exact Shapley calculation for a three-input forecasting system.
  Gray edges complete the coalition lattice; red edges are the marginal
  contributions from adding input 2. Averaging these contributions with the
  Shapley weights shown gives $\phi_2$. The same calculation applies to each
  other input.}
  \label{fig:exact-shapley-calculation}
\end{figure*}
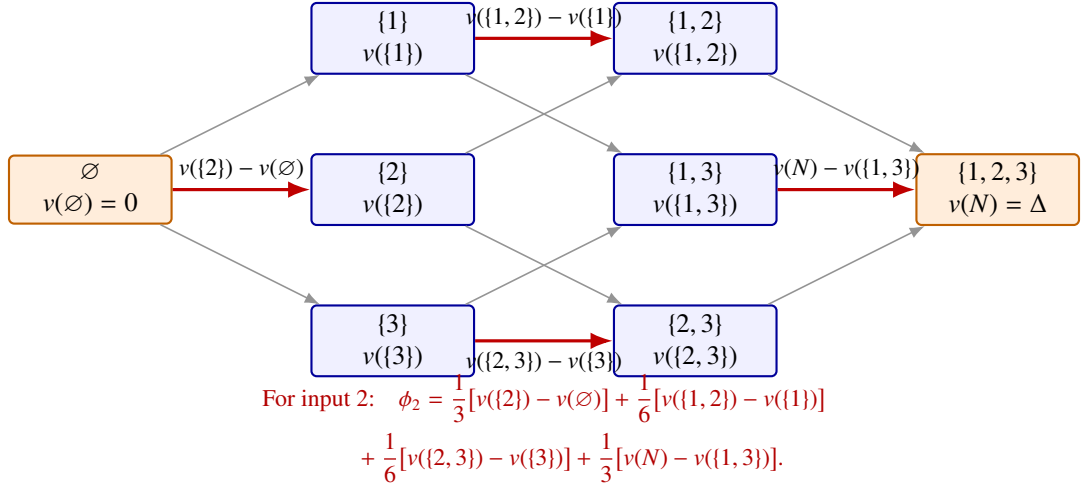

Figure~\ref{fig:exact-shapley-calculation} makes this calculation explicit for
three inputs. Each node is one coalition, and an edge from $S$ to
$S\cup\{i\}$ is the marginal contribution obtained by switching input $i$.
The highlighted edges are the four possible marginal contributions of input
2. Their weights depend only on the size of its predecessor coalition and sum
to one.

In other words, $\phi_i$ is the average value added when player $i$ switches
its input from $x_i^{(0)}$ to $x_i^{(1)}$. The average covers every coalition
that player $i$ could join and every ordering of the other players. Because
each permutation produces the exact decomposition in
Equation~\eqref{eq:walk-telescope}, the Shapley attributions satisfy the
efficiency property
\begin{align}
  \sum_{i=1}^{k}\phi_i
  &=\frac{1}{k!}
    \sum_{\pi\in\mathcal{S}_k}
    \sum_{i=1}^{k}C_i^{\pi} \notag\\
  &=\frac{1}{k!}
    \sum_{\pi\in\mathcal{S}_k}\Delta
   =\Delta.
  \label{eq:shapley-efficiency}
\end{align}
Thus, the Shapley value preserves the full forecast gap while removing the
dependence on any single, arbitrarily selected walk order.

\subsubsection{Computational cost}
The coalition formulation also clarifies the computational cost of an exact
Shapley calculation. For a fixed player $i$, the summation in
Equation~\eqref{eq:shapley-coalition} ranges over every subset of
$N\setminus\{i\}$. Because this set contains $k-1$ players, it has
\begin{equation}
  \left|2^{N\setminus\{i\}}\right|=2^{k-1}
  \label{eq:coalitions-per-player}
\end{equation}
possible coalitions. Each term compares $v(S)$ with $v(S\cup\{i\})$. Across
all players, these terms involve coalition values $v(T)$ for every subset
$T\subseteq N$. There are
\begin{equation}
  \left|2^N\right|=2^k
  \label{eq:total-coalitions}
\end{equation}
distinct coalitions, so an exact implementation that caches and reuses each
coalition value requires at most $2^k$ distinct evaluations of the forecasting
system. Since the baseline $F(\boldsymbol{x}^{\varnothing})$ and the second-run
forecast $F(\boldsymbol{x}^{N})$ are typically already available, only
$2^k-2$ additional hybrid runs may be needed. After these values are cached,
forming all player-level marginal differences requires $k2^{k-1}$ arithmetic
comparisons but no additional evaluations of $F$. Thus, the set formulation
reduces the naive $k!$ permutation enumeration to an exponential calculation,
but the cost can still become prohibitive when $k$ is large or when each
forecast run is expensive.

More explicitly, let $T_F$ denote the computational cost of one complete
evaluation of the forecasting system $F$, and let $T_A$ denote the cost of one
arithmetic operation used to form and aggregate a marginal contribution. If
the first- and second-run forecasts are already available, the total
computational work of the exact Shapley calculation is
\begin{equation}
  T_{\mathrm{Shapley}}(k)
  = (2^k-2)T_F
    +\mathcal{O}\!\left(k2^{k-1}T_A\right).
  \label{eq:shapley-total-cost}
\end{equation}
If the two endpoint forecasts must also be computed, the first term becomes
$2^kT_F$. In CCAR and CECL applications, a full evaluation of $F$ generally
dominates the arithmetic required to combine cached coalition values, so
Equation~\eqref{eq:shapley-total-cost} is typically well approximated by
\begin{equation}
  T_{\mathrm{Shapley}}(k)
  \approx (2^k-2)T_F.
  \label{eq:shapley-dominant-cost}
\end{equation}
This expression measures total computational work. Parallel execution can
reduce elapsed wall-clock time, but it does not reduce the total number of
forecasting-system evaluations. As the number of inputs $k$ grows, the
exponential number of coalition evaluations quickly becomes infeasible.

\subsubsection{Limitations}

Exact Shapley attribution requires every coalition-defined hybrid input to be
both executable and economically meaningful. This condition may fail when a
model version depends on a particular data schema, when portfolio records
cannot be aligned across launch points, or when an overlay is valid only under
a specified scenario. Such dependent inputs should be combined into a single
player or represented through a justified hierarchy. Even when all hybrids
are valid, the exponential run count limits exact enumeration to a relatively
small number of input blocks. Finally, the allocation is only as informative
as the player definition: grouping heterogeneous changes into one player can
hide material within-block interactions, whereas excessive granularity can
make the calculation operationally infeasible.

\subsection{Hierarchical/Nested Shapley}

\subsubsection{Methodology}

Inputs to a CCAR or CECL forecasting system often fall into economically
meaningful groups. As discussed in Section~\ref{sec:introduction}, the player
set $N$ may be partitioned into launch-point data, macroeconomic scenarios,
model specifications, business assumptions, and management overlays. A
hierarchical Shapley approach incorporates this structure instead of treating
all $k$ inputs as an unstructured set of interchangeable players. For a
two-level hierarchy, the resulting allocation is the Owen value for games with
a priori unions \cite{owen1977unions}.

Let
\begin{equation}
  \mathcal{G}=\{G_1,\ldots,G_m\}
  \label{eq:group-partition}
\end{equation}
be a partition of $N$, so that the groups are mutually disjoint and
$\bigcup_{g=1}^{m}G_g=N$. Write $k_g=|G_g|$. The hierarchy permits two kinds
of orderings: an outer permutation $\sigma$ of the $m$ groups and, for each
group $G_g$, an inner permutation $\tau_g$ of its $k_g$ players. The resulting
structured walk completes all players in one group before moving to the next
group.

For player $i\in G_g$, define its predecessor coalition under the structured
ordering $(\sigma,\boldsymbol{\tau})$ as
\begin{align}
  P_i^{\sigma,\boldsymbol{\tau}}
  \triangleq{}
  &\bigcup_{h:\,\sigma^{-1}(h)<\sigma^{-1}(g)}G_h \notag\\
  &{}\cup
  \left\{j\in G_g:
  \tau_g^{-1}(j)<\tau_g^{-1}(i)\right\},
  \label{eq:hierarchical-predecessors}
\end{align}
where $\boldsymbol{\tau}=(\tau_1,\ldots,\tau_m)$. Thus, all groups preceding
$G_g$ have fully played, no group following $G_g$ has played, and only the
players preceding $i$ within $G_g$ have played. The corresponding marginal
contribution is
\begin{equation}
  d_i^{\sigma,\boldsymbol{\tau}}
  \triangleq
  v\!\left(P_i^{\sigma,\boldsymbol{\tau}}\cup\{i\}\right)
  -v\!\left(P_i^{\sigma,\boldsymbol{\tau}}\right).
  \label{eq:hierarchical-marginal}
\end{equation}

The hierarchical, or Owen, attribution to player $i$ averages this marginal
contribution over every group ordering and every within-group ordering:
\begin{equation}
  \psi_i
  \triangleq
  \frac{1}{m!\prod_{h=1}^{m}k_h!}
  \sum_{\sigma\in\mathcal{S}_m}
  \sum_{\tau_1\in\mathcal{S}_{k_1}}\cdots
  \sum_{\tau_m\in\mathcal{S}_{k_m}}
  d_i^{\sigma,\boldsymbol{\tau}}.
  \label{eq:owen-value}
\end{equation}
Because every structured ordering is still a complete walk from
$\boldsymbol{x}^{(0)}$ to $\boldsymbol{x}^{(1)}$, the individual
attributions remain efficient:
\begin{equation}
  \sum_{i\in N}\psi_i=\Delta.
  \label{eq:owen-efficiency}
\end{equation}

The method also produces a coherent attribution at the group level. Define
the quotient game on the group index set $M=\{1,\ldots,m\}$ by
\begin{equation}
  V(A)
  \triangleq
  v\!\left(\bigcup_{g\in A}G_g\right),
  \qquad A\subseteq M.
  \label{eq:quotient-game}
\end{equation}
The Shapley value $\Phi_g$ of group $g$ in this outer game equals the sum of
the hierarchical attributions of its members,
\begin{equation}
  \Phi_g=\sum_{i\in G_g}\psi_i,
  \qquad
  \sum_{g=1}^{m}\Phi_g=\Delta.
  \label{eq:group-consistency}
\end{equation}
This property provides both a high-level attribution among major input blocks
and a detailed attribution among the inputs within each block, while ensuring
that the detailed results reconcile to the reported group totals.

For a deeper hierarchy, each group can be partitioned further. The same
construction is then applied recursively: the value assigned to a parent node
is distributed among its children, whose attributions sum to the parent total.
For example, the model group may be divided into PD, LGD, EAD, prepayment, and
rating-migration models, while the scenario group may be divided into general
economic conditions, the labor market, real estate, and interest rates.

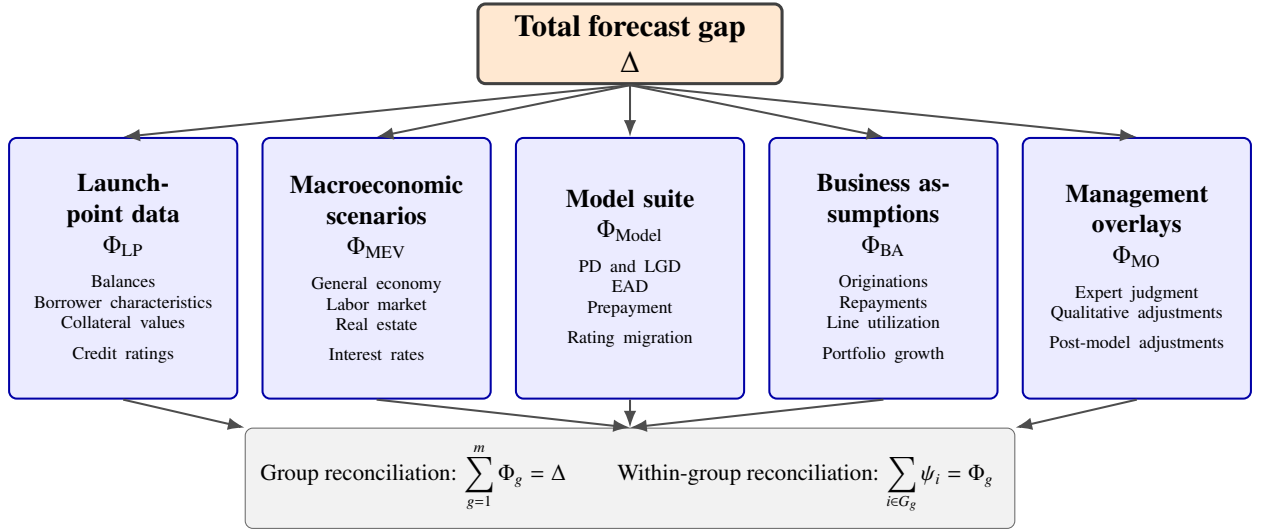
\begin{figure}[htbp]
  \centering
  \resizebox{\textwidth}{!}{%
  \begin{tikzpicture}[
    >=Latex,
    root/.style={
      draw=black!75,
      rounded corners=3pt,
      fill=orange!18,
      very thick,
      align=center,
      minimum width=4.2cm,
      minimum height=1.1cm,
      font=\large\bfseries
    },
    group/.style={
      draw=blue!65!black,
      rounded corners=3pt,
      fill=blue!8,
      thick,
      align=center,
      text width=2.8cm,
      minimum height=3.6cm,
      inner sep=5pt
    },
    identity/.style={
      draw=black!60,
      rounded corners=3pt,
      fill=gray!10,
      align=center,
      inner sep=6pt,
      font=\small
    },
    edge/.style={->,thick,draw=black!70}
  ]
    \node[root] (gap) at (0,0)
      {Total forecast gap\\$\Delta$};

    \node[group] (lp) at (-7.0,-3.1) {
      \textbf{Launch-point data}\\
      $\Phi_{\mathrm{LP}}$\\[2mm]
      \scriptsize
      Balances\\
      Borrower characteristics\\
      Collateral values\\
      Credit ratings
    };

    \node[group] (mev) at (-3.5,-3.1) {
      \textbf{Macroeconomic scenarios}\\
      $\Phi_{\mathrm{MEV}}$\\[2mm]
      \scriptsize
      General economy\\
      Labor market\\
      Real estate\\
      Interest rates
    };

    \node[group] (models) at (0,-3.1) {
      \textbf{Model suite}\\
      $\Phi_{\mathrm{Model}}$\\[2mm]
      \scriptsize
      PD and LGD\\
      EAD\\
      Prepayment\\
      Rating migration
    };

    \node[group] (business) at (3.5,-3.1) {
      \textbf{Business assumptions}\\
      $\Phi_{\mathrm{BA}}$\\[2mm]
      \scriptsize
      Originations\\
      Repayments\\
      Line utilization\\
      Portfolio growth
    };

    \node[group] (overlay) at (7.0,-3.1) {
      \textbf{Management overlays}\\
      $\Phi_{\mathrm{MO}}$\\[2mm]
      \scriptsize
      Expert judgment\\
      Qualitative adjustments\\
      Post-model adjustments
    };

    \draw[edge] (gap.south) -- (lp.north);
    \draw[edge] (gap.south) -- (mev.north);
    \draw[edge] (gap.south) -- (models.north);
    \draw[edge] (gap.south) -- (business.north);
    \draw[edge] (gap.south) -- (overlay.north);

    \node[identity] (reconcile) at (0,-6.0) {
      Group reconciliation: $\displaystyle\sum_{g=1}^{m}\Phi_g=\Delta$
      \qquad
      Within-group reconciliation:
      $\displaystyle\sum_{i\in G_g}\psi_i=\Phi_g$
    };

    \draw[edge] (lp.south) -- (reconcile.north west);
    \draw[edge] (mev.south) -- (reconcile.north);
    \draw[edge] (models.south) -- (reconcile.north);
    \draw[edge] (business.south) -- (reconcile.north);
    \draw[edge] (overlay.south) -- (reconcile.north east);
  \end{tikzpicture}%
  }
  \caption{Illustrative nested grouping for CCAR and CECL forecast-gap
  attribution. The total gap is first allocated among major input groups; each
  group attribution is then allocated among its nested inputs or subgroups.}
  \label{fig:nested-shapley-hierarchy}
\end{figure}

Figure~\ref{fig:nested-shapley-hierarchy} illustrates a possible hierarchy for
the CCAR and CECL attribution problem.

\subsubsection{Computational cost}

The hierarchy may also reduce the number of distinct hybrid forecast runs. A
two-level structured walk visits only coalitions consisting of several fully
activated groups and, at most, one partially activated group. With caching,
the number of distinct hierarchy-respecting coalition values is bounded by
\begin{equation}
  Q_{\mathcal{G}}
  \leq
  2^m
  +2^{m-1}\sum_{g=1}^{m}\left(2^{k_g}-2\right).
  \label{eq:hierarchical-coalition-count}
\end{equation}
Accordingly, if the endpoint forecasts are already available, the dominant
computational work is approximately
\begin{equation}
  T_{\mathrm{hier}}(\mathcal{G})
  \approx \left(Q_{\mathcal{G}}-2\right)T_F,
  \label{eq:hierarchical-cost}
\end{equation}
which can be substantially smaller than $(2^k-2)T_F$ when the hierarchy is
balanced. As with exact Shapley attribution, caching avoids repeated coalition
evaluations and parallel execution can reduce elapsed time without reducing
the total computational work.

\subsubsection{Limitations}

The computational reduction is obtained by changing the allocation rule: the
Owen value averages only over orders that respect the specified group
structure, not over all unrestricted player orders. Therefore, the resulting
attribution depends on the economic validity of the chosen hierarchy. Groups
should be defined before examining attribution results and should reflect
genuine operational, modeling, or governance relationships among the inputs.

It is important to distinguish the hierarchical attribution $\psi_i$ from the
unrestricted exact Shapley attribution $\phi_i$ in
Equation~\eqref{eq:shapley-permutation}. Both allocations are exact in the
efficiency sense: their player-level contributions sum to the full forecast
gap $\Delta$. In general, however,
\begin{equation}
  \psi_i \neq \phi_i,
  \label{eq:owen-versus-shapley}
\end{equation}
because the Shapley value averages over all $k!$ player permutations, whereas
the Owen value averages only over permutations in which the members of each
prespecified group remain together. The hierarchical attribution therefore
incorporates the chosen group structure into the allocation of interaction
effects. It coincides with the unrestricted Shapley value only in special
cases, such as when the grouping restrictions do not alter the relevant
marginal contributions. The computational saving therefore does not come from
calculating the same unrestricted Shapley values more efficiently. It comes
from adopting a different, group-structured allocation rule, which must be
justified by the economics and governance of the forecasting system.

\subsection{Integrated Gradients}

The exact Shapley and Owen-value methods above treat the forecasting system as
a black box and require only evaluations of $F$. When $F$ is differentiable
with respect to a continuous representation of its inputs, gradient-based path
methods provide a substantially cheaper alternative. Integrated Gradients was
introduced by Sundararajan et al.\ \cite{sundararajan2017integrated}.

\subsubsection{Methodology}

Integrated Gradients connects the first- and second-run inputs by the
straight-line path
\begin{equation}
  \boldsymbol{x}(\alpha)
  \triangleq
  \boldsymbol{x}^{(0)}
  +\alpha\left(\boldsymbol{x}^{(1)}-\boldsymbol{x}^{(0)}\right),
  \qquad \alpha\in[0,1].
  \label{eq:ig-path}
\end{equation}
The attribution to input $i$ is the accumulated sensitivity of the forecast
along this path, multiplied by the observed change in that input:
\begin{equation}
  \operatorname{IG}_i
  \triangleq
  \left(x_i^{(1)}-x_i^{(0)}\right)
  \int_0^1
  \frac{\partial F\!\left(\boldsymbol{x}(\alpha)\right)}{\partial x_i}
  \,\mathrm{d}\alpha.
  \label{eq:integrated-gradients}
\end{equation}
If $F$ is differentiable along the path, the fundamental theorem of calculus
gives the completeness property
\begin{align}
  \sum_{i=1}^{k}\operatorname{IG}_i
  &=\int_0^1
    \nabla F\!\left(\boldsymbol{x}(\alpha)\right)^{\mathsf T}
    \left(\boldsymbol{x}^{(1)}-\boldsymbol{x}^{(0)}\right)
    \,\mathrm{d}\alpha \notag\\
  &=F\!\left(\boldsymbol{x}^{(1)}\right)
    -F\!\left(\boldsymbol{x}^{(0)}\right)
   =\Delta.
  \label{eq:ig-completeness}
\end{align}
Integrated Gradients therefore provides an order-independent, gap-preserving
allocation for the selected continuous path. It remains path dependent,
however, because a different interpolation path may produce a different
allocation.

\begin{figure*}[htbp]
  \centering
  \resizebox{\textwidth}{!}{%
  \begin{tikzpicture}[
    >=Latex,
    endpoint/.style={
      draw=orange!75!black,
      rounded corners=3pt,
      fill=orange!14,
      thick,
      align=center,
      minimum width=3.0cm,
      minimum height=1.15cm,
      inner sep=5pt
    },
    pathpoint/.style={
      draw=blue!60!black,
      rounded corners=3pt,
      fill=blue!6,
      thick,
      align=center,
      minimum width=3.2cm,
      minimum height=1.15cm,
      inner sep=5pt
    },
    aggregate/.style={
      draw=green!45!black,
      rounded corners=3pt,
      fill=green!8,
      very thick,
      align=center,
      minimum width=5.2cm,
      minimum height=1.55cm,
      inner sep=7pt
    },
    patharrow/.style={->,thick,draw=black!65},
    collect/.style={->,thick,draw=green!45!black}
  ]
    \node[endpoint] (x0) at (0,0) {
      \textbf{First-run input}\\
      $\boldsymbol{x}^{(0)}$};
    \node[pathpoint] (a1) at (4.4,0) {
      $\boldsymbol{x}(\alpha_1)$\\
      $\nabla F(\boldsymbol{x}(\alpha_1))$};
    \node[pathpoint] (ar) at (8.8,0) {
      $\boldsymbol{x}(\alpha_r)$\\
      $\nabla F(\boldsymbol{x}(\alpha_r))$};
    \node[pathpoint] (aM) at (13.2,0) {
      $\boldsymbol{x}(\alpha_M)$\\
      $\nabla F(\boldsymbol{x}(\alpha_M))$};
    \node[endpoint] (x1) at (17.6,0) {
      \textbf{Second-run input}\\
      $\boldsymbol{x}^{(1)}$};

    \draw[patharrow] (x0) -- node[above,font=\small] {$\alpha_1$} (a1);
    \draw[patharrow] (a1) -- node[above,font=\small] {$\cdots$} (ar);
    \draw[patharrow] (ar) -- node[above,font=\small] {$\cdots$} (aM);
    \draw[patharrow] (aM) -- (x1);

    \node[aggregate] (ig) at (8.8,-3.0) {
      \textbf{Average gradients and scale by the input change}\\[1mm]
      $\displaystyle
       \widehat{\operatorname{IG}}_i
       =(x_i^{(1)}-x_i^{(0)})
       \frac{1}{M}\sum_{r=1}^{M}
       \frac{\partial F(\boldsymbol{x}(\alpha_r))}{\partial x_i}$};

    \draw[collect] (a1.south) -- (ig.north west);
    \draw[collect] (ar.south) -- (ig.north);
    \draw[collect] (aM.south) -- (ig.north east);
  \end{tikzpicture}%
  }
  \caption{Integrated Gradients calculation. Points are selected along the
  straight-line path from the first-run input to the second-run input. The
  forecasting-system gradient is evaluated at each point, averaged, and
  multiplied componentwise by the change between the two runs. With exact
  integration, the input attributions sum to the forecast gap $\Delta$.}
  \label{fig:ig-calculation}
\end{figure*}
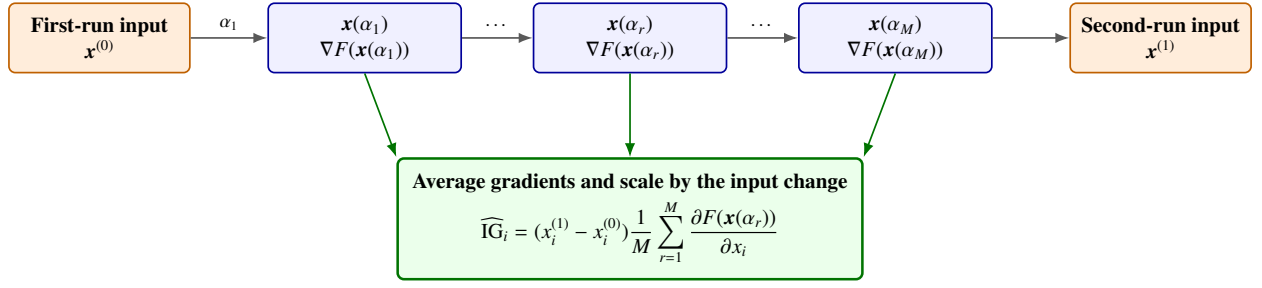

In practice, the integral is approximated using $M$ points
$0<\alpha_1<\cdots<\alpha_M\leq1$:
\begin{equation}
  \widehat{\operatorname{IG}}_i
  =\left(x_i^{(1)}-x_i^{(0)}\right)
   \frac{1}{M}\sum_{r=1}^{M}
   \frac{\partial F\!\left(\boldsymbol{x}(\alpha_r)\right)}{\partial x_i}.
  \label{eq:ig-estimator}
\end{equation}
Figure~\ref{fig:ig-calculation} illustrates this numerical calculation. The
same interpolation points are used for every input, a gradient is evaluated at
each point, and the resulting average sensitivity is scaled by the observed
input change.

\subsubsection{Computational cost}

Let $T_{\nabla F}$ denote the cost of evaluating the gradient of $F$. The
dominant computational work is approximately
\begin{equation}
  T_{\mathrm{IG}}(M)\approx M T_{\nabla F},
  \label{eq:ig-cost}
\end{equation}
which does not grow exponentially with $k$. The numerical completeness error
\begin{equation}
  \varepsilon_{\mathrm{IG}}
  \triangleq
  \Delta-\sum_{i=1}^{k}\widehat{\operatorname{IG}}_i
  \label{eq:ig-error}
\end{equation}
can be monitored and reduced by increasing $M$ or using a more accurate
quadrature rule.

\subsubsection{Limitations}

Integrated Gradients requires both forecast runs to be represented as vectors
of differentiable inputs and requires the straight-line interpolation between
them to be meaningful. These conditions are restrictive in CCAR and CECL
systems. A launch-point file may contain different loans across runs, a model
update may replace code or architecture, and a qualitative overlay may be a
discrete governance decision. Interpolating between such objects can create
invalid intermediate states, while production components may not expose the
derivatives needed by the method.

Even continuous quantities such as balances, MEV paths, and loan attributes
are often transformed into discrete buckets. The resulting step functions
have derivatives that are zero almost everywhere and undefined at bucket
boundaries. In addition, Integrated Gradients is a straight-path
Aumann--Shapley allocation rather than the exact discrete Shapley value, and a
different valid interpolation path may produce a different attribution.
Discrete or non-differentiable changes should therefore be retained as
Shapley or Owen players, potentially within a hybrid attribution framework.

\subsection{Gradient SHAP and Expected Gradients}

Gradient SHAP combines the SHAP framework \cite{lundberg2017shap} with the
expected-gradients construction \cite{erion2021expected}. It extends
Integrated Gradients by averaging over reference inputs and locations on the
paths from those references to the second-run input.

\subsubsection{Methodology}

\begin{figure*}[htbp]
  \centering
  \resizebox{\textwidth}{!}{%
  \begin{tikzpicture}[
    >=Latex,
    baseline/.style={
      draw=orange!75!black,
      rounded corners=3pt,
      fill=orange!14,
      thick,
      align=center,
      minimum width=2.8cm,
      minimum height=1.05cm,
      inner sep=5pt
    },
    sample/.style={
      draw=blue!60!black,
      rounded corners=3pt,
      fill=blue!6,
      thick,
      align=center,
      minimum width=4.6cm,
      minimum height=1.15cm,
      inner sep=5pt
    },
    target/.style={baseline,minimum width=2.8cm},
    average/.style={
      draw=green!45!black,
      rounded corners=3pt,
      fill=green!8,
      very thick,
      align=center,
      minimum width=6.0cm,
      minimum height=1.55cm,
      inner sep=7pt
    },
    patharrow/.style={->,thick,draw=black!65},
    collect/.style={->,thick,draw=green!45!black},
    rowlabel/.style={font=\small\bfseries,anchor=east}
  ]
    \node[rowlabel] at (-0.3,2.8) {Draw $r=1$:};
    \node[baseline] (b1) at (1.4,2.8)
      {$\boldsymbol{b}^{(1)}\sim\mathcal{B}$};
    \node[sample] (z1) at (7.0,2.8)
      {$\boldsymbol{z}^{(1)}=\boldsymbol{b}^{(1)}
        +\alpha_1(\boldsymbol{x}^{(1)}-\boldsymbol{b}^{(1)})$\\
       $\nabla F(\boldsymbol{z}^{(1)})$};
    \node[target] (t1) at (12.6,2.8) {$\boldsymbol{x}^{(1)}$};

    \node[rowlabel] at (-0.3,0.5) {Draw $r=2$:};
    \node[baseline] (b2) at (1.4,0.5)
      {$\boldsymbol{b}^{(2)}\sim\mathcal{B}$};
    \node[sample] (z2) at (7.0,0.5)
      {$\boldsymbol{z}^{(2)}=\boldsymbol{b}^{(2)}
        +\alpha_2(\boldsymbol{x}^{(1)}-\boldsymbol{b}^{(2)})$\\
       $\nabla F(\boldsymbol{z}^{(2)})$};
    \node[target] (t2) at (12.6,0.5) {$\boldsymbol{x}^{(1)}$};

    \node at (7.0,-1.0) {$\vdots$};

    \node[rowlabel] at (-0.3,-2.5) {Draw $r=R$:};
    \node[baseline] (bR) at (1.4,-2.5)
      {$\boldsymbol{b}^{(R)}\sim\mathcal{B}$};
    \node[sample] (zR) at (7.0,-2.5)
      {$\boldsymbol{z}^{(R)}=\boldsymbol{b}^{(R)}
        +\alpha_R(\boldsymbol{x}^{(1)}-\boldsymbol{b}^{(R)})$\\
       $\nabla F(\boldsymbol{z}^{(R)})$};
    \node[target] (tR) at (12.6,-2.5) {$\boldsymbol{x}^{(1)}$};

    \draw[patharrow] (b1) -- node[above,font=\small] {$\alpha_1$} (z1);
    \draw[patharrow] (z1) -- (t1);
    \draw[patharrow] (b2) -- node[above,font=\small] {$\alpha_2$} (z2);
    \draw[patharrow] (z2) -- (t2);
    \draw[patharrow] (bR) -- node[above,font=\small] {$\alpha_R$} (zR);
    \draw[patharrow] (zR) -- (tR);

    \node[average] (avg) at (18.1,0.2) {
      \textbf{Average scaled gradients by input}\\[1mm]
      $\displaystyle
       \widehat{\operatorname{GS}}_i
       =\frac{1}{R}\sum_{r=1}^{R}
       (x_i^{(1)}-b_i^{(r)})
       \frac{\partial F(\boldsymbol{z}^{(r)})}{\partial x_i}$};

    \draw[collect] (t1.east) -- (avg.north west);
    \draw[collect] (t2.east) -- (avg.west);
    \draw[collect] (tR.east) -- (avg.south west);
  \end{tikzpicture}%
  }
  \caption{Gradient SHAP calculation. Each Monte Carlo draw samples a baseline
  $\boldsymbol{b}^{(r)}$ and a path location $\alpha_r$, evaluates the gradient
  at the resulting interpolation point, and scales each gradient component by
  the corresponding change from the sampled baseline to the second-run input.
  Averaging the scaled gradients gives the Gradient SHAP attribution.}
  \label{fig:gradient-shap-calculation}
\end{figure*}
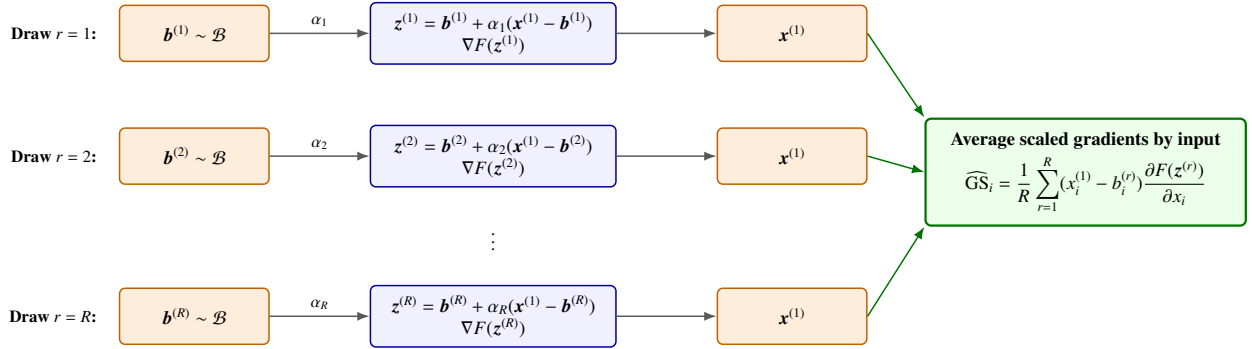

Integrated Gradients uses one fixed baseline $\boldsymbol{x}^{(0)}$ and one
deterministic path. Gradient SHAP instead samples a reference input
$\boldsymbol{b}$ from a baseline distribution $\mathcal{B}$ and samples a
random point on the path from $\boldsymbol{b}$ to
$\boldsymbol{x}^{(1)}$. Its population attribution can be written as
\begin{equation}
  \operatorname{GS}_i
  \triangleq
  \mathbb{E}_{\substack{\boldsymbol{b}\sim\mathcal{B}\\
                         \alpha\sim\operatorname{Unif}(0,1)}}
  \left[
    \left(x_i^{(1)}-b_i\right)
    \frac{\partial
      F\!\left(\boldsymbol{b}
      +\alpha(\boldsymbol{x}^{(1)}-\boldsymbol{b})\right)}
    {\partial x_i}
  \right].
  \label{eq:gradient-shap}
\end{equation}
With $R$ independent draws
$(\boldsymbol{b}^{(r)},\alpha_r)$, a Monte Carlo estimator is
\begin{equation}
  \widehat{\operatorname{GS}}_i
  =\frac{1}{R}\sum_{r=1}^{R}
  \left(x_i^{(1)}-b_i^{(r)}\right)
  \frac{\partial
    F\!\left(\boldsymbol{b}^{(r)}
    +\alpha_r(\boldsymbol{x}^{(1)}-\boldsymbol{b}^{(r)})\right)}
  {\partial x_i}.
  \label{eq:gradient-shap-estimator}
\end{equation}
Figure~\ref{fig:gradient-shap-calculation} illustrates the Monte Carlo
calculation. Each draw selects both a baseline and a point on the corresponding
straight-line path to the second-run input. The scaled gradients from these
draws are then averaged by input.

Under exact integration, these attributions satisfy completeness relative to
the expected baseline forecast:
\begin{equation}
  \sum_{i=1}^{k}\operatorname{GS}_i
  =F\!\left(\boldsymbol{x}^{(1)}\right)
   -\mathbb{E}_{\boldsymbol{b}\sim\mathcal{B}}
    \left[F(\boldsymbol{b})\right].
  \label{eq:gradient-shap-completeness}
\end{equation}
If $\mathcal{B}$ places all its mass on the first-run input
$\boldsymbol{x}^{(0)}$, Gradient SHAP reduces to a stochastic estimator of
Integrated Gradients and the right-hand side of
Equation~\eqref{eq:gradient-shap-completeness} becomes $\Delta$. A broader
baseline distribution can represent multiple plausible reference runs or
uncertainty around the first run, but the resulting attribution then explains
the gap relative to an expected reference forecast rather than the specific
two-run gap in Equation~\eqref{eq:total-gap}.

\subsubsection{Computational cost}

With $R$ sampled baseline--path-point pairs, the dominant computational cost is
\begin{equation}
  T_{\mathrm{GS}}(R)\approx R T_{\nabla F}.
  \label{eq:gradient-shap-cost}
\end{equation}

This cost grows linearly with the number of Monte Carlo draws rather than
exponentially with the number of inputs. Each gradient evaluation contributes
to every input attribution, although additional draws may be needed when the
baseline distribution or local gradients are highly variable. Fixed random
seeds, Monte Carlo standard errors, and the completeness residual provide
practical numerical and reproducibility diagnostics.

\subsubsection{Limitations}

To use Gradient SHAP, the target and sampled baselines must be represented as
vectors of differentiable inputs. Continuous quantities such as balances, MEV
paths, loan attributes, model parameters, and scalar business assumptions can
be interpolated directly. Feature-level attributions may then be aggregated
within the groups in Figure~\ref{fig:nested-shapley-hierarchy} to obtain
portfolio-data, scenario, model, business-assumption, and overlay
contributions.

Important limitations arise because many production inputs are not naturally
continuous. A launch-point file may contain different loans across runs; a
model update may replace code or model architecture; and a qualitative overlay
may be a discrete governance decision. Straight-line interpolation between
such objects can produce economically invalid intermediate states, and a
production forecasting system may not expose derivatives through every
component. In those cases, one must use a differentiable parameterization or
surrogate, keep the affected block as a discrete player, or combine gradient
attribution for continuous inputs with Shapley or Owen attribution for discrete
blocks.

Even apparently continuous quantities such as balances, MEV paths, and loan
attributes are often transformed into discrete buckets in risk models. Such
transformations are step functions whose derivatives are zero almost
everywhere and undefined at the bucket boundaries, thereby limiting the
usefulness of gradient-based attribution methods.

Gradient SHAP is not generally identical to the exact discrete Shapley value
$\phi_i$ in Equation~\eqref{eq:shapley-coalition}. Its result depends on the
selected baseline distribution as well as the validity of the interpolation
paths. Unless the distribution is concentrated on
$\boldsymbol{x}^{(0)}$, the attributions do not reconcile to the specific
two-run gap $\Delta$, but instead to a gap measured from the expected baseline
forecast. The baseline population and sampling design must therefore have a
clear economic interpretation and be retained for reproducibility and
governance.

\subsection{Permutation SHAP}

\subsubsection{Methodology}

Exact Shapley attribution becomes costly because it requires the evaluation of
all $2^k$ hybrid coalitions. Permutation SHAP, based on permutation sampling
\cite{castro2009polynomial}, directly approximates
Equation~\eqref{eq:shapley-permutation} by Monte Carlo. Instead of averaging
over all $k!$ walk orders, it samples a manageable number and averages the
resulting contributions. Like the exact method, it treats the forecasting
system as a black box. It therefore requires neither differentiability nor a
potentially artificial interpolation between the two runs.

\begin{figure*}[htbp]
  \centering
  \resizebox{\textwidth}{!}{%
  \begin{tikzpicture}[
    >=Latex,
    endpoint/.style={
      draw=orange!75!black,
      rounded corners=3pt,
      fill=orange!14,
      thick,
      align=center,
      minimum width=2.8cm,
      minimum height=1.05cm,
      inner sep=5pt
    },
    hybrid/.style={
      draw=blue!60!black,
      rounded corners=3pt,
      fill=blue!6,
      thick,
      align=center,
      minimum width=2.35cm,
      minimum height=1.05cm,
      inner sep=5pt
    },
    average/.style={
      draw=green!45!black,
      rounded corners=3pt,
      fill=green!8,
      very thick,
      align=center,
      minimum width=4.1cm,
      minimum height=1.4cm,
      inner sep=7pt
    },
    walkarrow/.style={->,thick,draw=black!65},
    collect/.style={->,thick,draw=green!45!black},
    rowlabel/.style={font=\small\bfseries,anchor=east}
  ]
    \node[rowlabel] at (-0.3, 3.0) {Sample $b=1$:};
    \node[endpoint] (s01) at (1.4,3.0)
      {$\boldsymbol{x}^{(0)}$\\$F(\boldsymbol{x}^{(0)})$};
    \node[hybrid] (h11) at (5.1,3.0)
      {$\boldsymbol{x}^{\pi^{(1)},1}$\\$F(\boldsymbol{x}^{\pi^{(1)},1})$};
    \node[hybrid] (hd1) at (8.8,3.0) {$\cdots$};
    \node[endpoint] (sk1) at (12.5,3.0)
      {$\boldsymbol{x}^{(1)}$\\$F(\boldsymbol{x}^{(1)})$};

    \node[rowlabel] at (-0.3, 0.8) {Sample $b=2$:};
    \node[endpoint] (s02) at (1.4,0.8)
      {$\boldsymbol{x}^{(0)}$\\$F(\boldsymbol{x}^{(0)})$};
    \node[hybrid] (h12) at (5.1,0.8)
      {$\boldsymbol{x}^{\pi^{(2)},1}$\\$F(\boldsymbol{x}^{\pi^{(2)},1})$};
    \node[hybrid] (hd2) at (8.8,0.8) {$\cdots$};
    \node[endpoint] (sk2) at (12.5,0.8)
      {$\boldsymbol{x}^{(1)}$\\$F(\boldsymbol{x}^{(1)})$};

    \node at (6.9,-0.65) {$\vdots$};

    \node[rowlabel] at (-0.3,-2.1) {Sample $b=B$:};
    \node[endpoint] (s0B) at (1.4,-2.1)
      {$\boldsymbol{x}^{(0)}$\\$F(\boldsymbol{x}^{(0)})$};
    \node[hybrid] (h1B) at (5.1,-2.1)
      {$\boldsymbol{x}^{\pi^{(B)},1}$\\$F(\boldsymbol{x}^{\pi^{(B)},1})$};
    \node[hybrid] (hdB) at (8.8,-2.1) {$\cdots$};
    \node[endpoint] (skB) at (12.5,-2.1)
      {$\boldsymbol{x}^{(1)}$\\$F(\boldsymbol{x}^{(1)})$};

    \draw[walkarrow] (s01) -- node[above,font=\small]
      {$C_{\pi^{(1)}_1}^{\pi^{(1)}}$} (h11);
    \draw[walkarrow] (h11) -- node[above,font=\small]
      {$C_{\pi^{(1)}_2}^{\pi^{(1)}},\ldots$} (hd1);
    \draw[walkarrow] (hd1) -- node[above,font=\small]
      {$\ldots,C_{\pi^{(1)}_k}^{\pi^{(1)}}$} (sk1);

    \draw[walkarrow] (s02) -- node[above,font=\small]
      {$C_{\pi^{(2)}_1}^{\pi^{(2)}}$} (h12);
    \draw[walkarrow] (h12) -- node[above,font=\small]
      {$C_{\pi^{(2)}_2}^{\pi^{(2)}},\ldots$} (hd2);
    \draw[walkarrow] (hd2) -- node[above,font=\small]
      {$\ldots,C_{\pi^{(2)}_k}^{\pi^{(2)}}$} (sk2);

    \draw[walkarrow] (s0B) -- node[above,font=\small]
      {$C_{\pi^{(B)}_1}^{\pi^{(B)}}$} (h1B);
    \draw[walkarrow] (h1B) -- node[above,font=\small]
      {$C_{\pi^{(B)}_2}^{\pi^{(B)}},\ldots$} (hdB);
    \draw[walkarrow] (hdB) -- node[above,font=\small]
      {$\ldots,C_{\pi^{(B)}_k}^{\pi^{(B)}}$} (skB);

    \node[average] (avg) at (17.0,0.45) {
      \textbf{Average by input}\\[1mm]
      $\displaystyle
       \widehat{\phi}_i^{\mathrm{perm}}
       =\frac{1}{B}\sum_{b=1}^{B}C_i^{\pi^{(b)}}$\\
      $i=1,\ldots,k$};

    \draw[collect] (sk1.east) -- (avg.north west);
    \draw[collect] (sk2.east) -- (avg.west);
    \draw[collect] (skB.east) -- (avg.south west);
  \end{tikzpicture}%
  }
  \caption{Permutation SHAP calculation. Each sampled permutation defines a
  complete walk from the first-run inputs to the second-run inputs. The
  contribution of each input is recorded along every walk and averaged across
  the $B$ sampled orders. Each row reconciles to $\Delta$, so the averaged
  attributions also reconcile exactly to the total forecast gap.}
  \label{fig:permutation-shap-calculation}
\end{figure*}
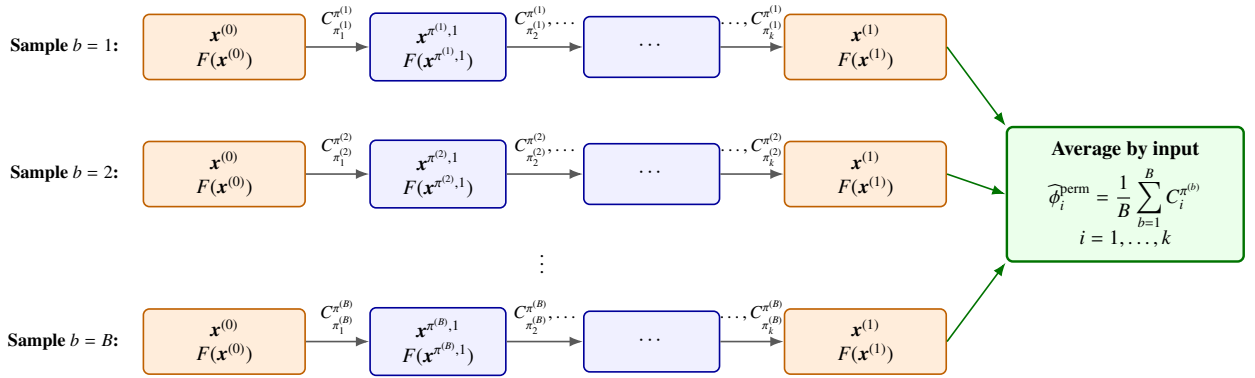

Let $\pi^{(1)},\ldots,\pi^{(B)}$ be independent permutations sampled uniformly
from $\mathcal{S}_k$. For each sampled order, the forecasting system is run
along the sequence of hybrid inputs
$\boldsymbol{x}^{\pi^{(b)},0},\ldots,
\boldsymbol{x}^{\pi^{(b)},k}$ defined in
Equation~\eqref{eq:hybrid-input}. The contribution of input $i$ on the $b$th
walk is
\begin{equation}
  C_i^{\pi^{(b)}}
  =F\!\left(\boldsymbol{x}^{\pi^{(b)},j_b(i)}\right)
   -F\!\left(\boldsymbol{x}^{\pi^{(b)},j_b(i)-1}\right),
  \qquad
  j_b(i)\triangleq (\pi^{(b)})^{-1}(i),
  \label{eq:permutation-shap-walk-contribution}
\end{equation}
where $j_b(i)$ is the step at which input $i$ is changed. The Permutation SHAP
estimator is the sample average
\begin{equation}
  \widehat{\phi}_i^{\mathrm{perm}}
  \triangleq
  \frac{1}{B}\sum_{b=1}^{B}C_i^{\pi^{(b)}},
  \qquad i=1,\ldots,k.
  \label{eq:permutation-shap-estimator}
\end{equation}
Figure~\ref{fig:permutation-shap-calculation} illustrates the calculation.
Each row is one randomly sampled walk from the first run to the second run.
The marginal contribution recorded when input $i$ is switched is then averaged
vertically across the sampled walks.

Because the permutations are sampled uniformly,
$\mathbb{E}[\widehat{\phi}_i^{\mathrm{perm}}]=\phi_i$; thus, the estimator is
unbiased for the unrestricted Shapley value. Moreover, every sampled walk
telescopes to $\Delta$. Consequently, using the same sampled walks for all
inputs gives exact sample-level reconciliation,
\begin{equation}
  \sum_{i=1}^{k}\widehat{\phi}_i^{\mathrm{perm}}=\Delta,
  \label{eq:permutation-shap-efficiency}
\end{equation}
apart from numerical or production-run error. The individual allocations are
random estimates, but the total attributed amount is not.

In this setting, each player may represent a major input block,
such as launch-point data, scenario MEVs, models, business assumptions, or
management overlays, or a more granular component within one of those blocks.
For a sampled permutation, one begins with the first-run configuration and
replaces the inputs in the sampled order, recording the forecast after every
replacement. Consecutive forecast differences are assigned to the input just
changed. Repeating the procedure across sampled orders and averaging by input
estimates the order-independent allocation of the CCAR or CECL forecast gap. A
fixed random seed and a retained inventory of the sampled orders make the
analysis reproducible and auditable.

\subsubsection{Computational cost}

If the two endpoint forecasts are already available, a single sampled walk
requires at most $k-1$ additional hybrid evaluations. Without reuse across
walks, the dominant computational work is therefore approximately
\begin{equation}
  T_{\mathrm{perm}}(B,k)
  \approx B(k-1)T_F,
  \label{eq:permutation-shap-cost}
\end{equation}
which grows linearly in the number of sampled permutations rather than
exponentially in $k$. Caching a hybrid coalition encountered in more than one
walk can reduce the actual number of evaluations. The walks can also be run in
parallel, subject to the operational capacity of the forecasting platform.

\subsubsection{Limitations}

Sampling uncertainty should be reported because a finite collection of walks
does not remove order effects completely. For each input, an estimated Monte
Carlo standard error is
\begin{equation}
  \widehat{\operatorname{se}}_i
  =\sqrt{\frac{1}{B(B-1)}
    \sum_{b=1}^{B}
    \left(C_i^{\pi^{(b)}}-
    \widehat{\phi}_i^{\mathrm{perm}}\right)^2}.
  \label{eq:permutation-shap-standard-error}
\end{equation}
Sampling may be continued until these standard errors, or corresponding
confidence-interval widths, fall below prespecified materiality thresholds.
Using a sampled order together with its reverse is a simple variance-reduction
device because it exposes each input to complementary predecessor coalitions.
Stratifying the samples by the position of each input can further improve
coverage of early, middle, and late walk positions.

Permutation SHAP approximates the unrestricted exact Shapley value without
changing the underlying allocation rule, unlike a hierarchy-restricted Owen
value. Its main limitation is that a stable estimate may still require many
expensive hybrid runs when interactions are strong. In addition, every hybrid
configuration must be operationally executable and economically interpretable.
These requirements are particularly important when launch-point populations,
model versions, or overlays cannot be switched independently. In such cases,
the affected inputs should be combined into a single player or handled through
an economically justified hierarchy before permutation sampling is applied.

\subsection{Kernel SHAP}

\subsubsection{Methodology}

Kernel SHAP is another model-agnostic approximation to the exact Shapley value
\cite{lundberg2017shap}. Instead of sampling complete walks, it samples
coalitions, evaluates the corresponding hybrid forecasts, and estimates the
input attributions through a specially weighted linear regression. The method
is therefore applicable when the forecasting system is available only as a
black box and its derivatives are unavailable.

For a coalition $S\subseteq N$, define the binary indicator vector
$\boldsymbol{z}^{S}\in\{0,1\}^{k}$ by $z_i^{S}=1$ if $i\in S$ and
$z_i^{S}=0$ otherwise. Kernel SHAP approximates the coalition-value function
by the additive explanation model
\begin{equation}
  g(\boldsymbol{z}^{S})
  \triangleq
  \phi_0+\sum_{i=1}^{k}\phi_i z_i^{S},
  \label{eq:kernel-shap-additive-model}
\end{equation}
where $\phi_i$ is the attribution assigned to input $i$. When the regression
response is the forecast level $F(\boldsymbol{x}^{S})$, the intercept is fixed
at $\phi_0=F(\boldsymbol{x}^{(0)})$. Equivalently, using $v(S)$ from
Equation~\eqref{eq:coalition-value} as the response fixes $\phi_0=0$.

Suppose $M$ coalitions $S_1,\ldots,S_M$ are sampled and their hybrid forecasts
are evaluated. The Kernel SHAP estimate solves
\begin{equation}
  \widehat{\boldsymbol{\phi}}^{\mathrm{ker}}
  \triangleq
  \underset{\boldsymbol{\phi}}{\operatorname{argmin}}
  \sum_{m=1}^{M}
  \pi_{\mathrm{SHAP}}(S_m)
  \left[
    v(S_m)-\sum_{i=1}^{k}\phi_i z_i^{S_m}
  \right]^2,
  \label{eq:kernel-shap-regression}
\end{equation}
subject to the efficiency constraint
\begin{equation}
  \sum_{i=1}^{k}\phi_i=\Delta.
  \label{eq:kernel-shap-efficiency}
\end{equation}
For a nonempty, non-full coalition, the Shapley kernel weight is
\begin{equation}
  \pi_{\mathrm{SHAP}}(S)
  \triangleq
  \frac{k-1}
  {\binom{k}{|S|}\,|S|\,(k-|S|)},
  \qquad 1\leq |S|\leq k-1.
  \label{eq:kernel-shap-weight}
\end{equation}
The empty and full coalitions are imposed as exact endpoint constraints rather
than assigned finite weights. The kernel gives relatively high total influence
to coalitions near the two endpoints, where the incremental effect of adding
or removing an input is especially informative. If every coalition is
evaluated and the regression is solved without regularization, its coefficients
equal the exact Shapley values. With a sampled subset of coalitions, the fitted
coefficients provide an approximation.

To apply Kernel SHAP to two CCAR or CECL runs, the analyst first defines the
players, such as launch-point data, scenario MEVs, model versions, business
assumptions, and management overlays. For each sampled coalition $S_m$, inputs
in $S_m$ are taken from the second run and the remaining inputs are taken from
the first run, exactly as in Equation~\eqref{eq:coalition-input}. The resulting
hybrid configuration is run through the forecasting system to obtain
$v(S_m)$. The weighted regression in
Equation~\eqref{eq:kernel-shap-regression} then allocates the total forecast gap
among the input changes. Coalition sampling should cover a range of coalition
sizes, with probabilities aligned to the Shapley kernel; complementary pairs
$S$ and $N\setminus S$ can be sampled together to improve balance. A fixed
sample, retained hybrid-run specifications, and the fitted regression design
make the analysis reproducible and auditable.

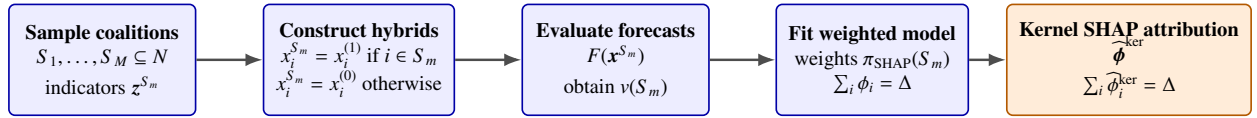
\begin{figure}[htbp]
  \centering
  \resizebox{\textwidth}{!}{%
  \begin{tikzpicture}[
    >=Latex,
    flowbox/.style={
      draw=blue!65!black,
      rounded corners=3pt,
      fill=blue!7,
      thick,
      align=center,
      minimum width=3.2cm,
      minimum height=1.9cm,
      inner sep=6pt
    },
    endpoint/.style={flowbox,fill=orange!15,draw=orange!70!black},
    flow/.style={->,very thick,draw=black!70}
  ]
    \node[flowbox] (sample) at (0,0) {
      \textbf{Sample coalitions}\\
      $S_1,\ldots,S_M\subseteq N$\\[1mm]
      indicators $\boldsymbol{z}^{S_m}$
    };

    \node[flowbox] (hybrid) at (4.4,0) {
      \textbf{Construct hybrids}\\
      $x_i^{S_m}=x_i^{(1)}$ if $i\in S_m$\\
      $x_i^{S_m}=x_i^{(0)}$ otherwise
    };

    \node[flowbox] (evaluate) at (8.8,0) {
      \textbf{Evaluate forecasts}\\
      $F(\boldsymbol{x}^{S_m})$\\[1mm]
      obtain $v(S_m)$
    };

    \node[flowbox] (regress) at (13.2,0) {
      \textbf{Fit weighted model}\\
      weights $\pi_{\mathrm{SHAP}}(S_m)$\\
      $\sum_i\phi_i=\Delta$
    };

    \node[endpoint] (output) at (17.6,0) {
      \textbf{Kernel SHAP attribution}\\
      $\widehat{\boldsymbol{\phi}}^{\mathrm{ker}}$\\[1mm]
      $\sum_i\widehat{\phi}_i^{\mathrm{ker}}=\Delta$
    };

    \draw[flow] (sample) -- (hybrid);
    \draw[flow] (hybrid) -- (evaluate);
    \draw[flow] (evaluate) -- (regress);
    \draw[flow] (regress) -- (output);
  \end{tikzpicture}%
  }
  \caption{Kernel SHAP workflow for attributing the gap between two forecast
  runs. Sampled coalitions define hybrid inputs, the forecasting system
  supplies coalition values, and a Shapley-kernel-weighted regression subject
  to efficiency produces order-independent input attributions.}
  \label{fig:kernel-shap-workflow}
\end{figure}

Figure~\ref{fig:kernel-shap-workflow} summarizes this procedure. Unlike a
sampled walk, each coalition evaluation is independent of the others; the
evaluated coalition values are combined only at the final weighted-regression
stage.

\subsubsection{Computational cost}

If the two endpoint forecasts are already available, the dominant cost of
$M$ distinct sampled coalitions is approximately
\begin{equation}
  T_{\mathrm{kernel}}(M)\approx M T_F,
  \label{eq:kernel-shap-cost}
\end{equation}
with further savings when coalition forecasts are cached or evaluated in
parallel. Kernel SHAP can thus use a flexible evaluation budget and does not
require the $k-1$ linked hybrid runs needed to complete each sampled
permutation.

\subsubsection{Limitations}

A sufficiently large and well-balanced coalition sample is needed for a stable
regression, especially when $k$ is large or input effects interact strongly.
Repeated coalition samples or bootstrap resampling can be used to assess
numerical and sampling variability.

The method has the same operational limitation as exact and Permutation
SHAP: every sampled hybrid configuration must be executable and economically
meaningful. For example, a model version may require a particular data schema,
or a management overlay may have been approved only for a particular scenario.
Inputs with such dependencies should be combined into one player or represented
through an economically justified hierarchy. In addition, aggressive feature
selection or penalized regression may stabilize an underdetermined fit but can
change the allocation and prevent it from representing the unrestricted
Shapley value. For a small number of major input blocks, exact enumeration may
therefore remain preferable; Kernel SHAP is most useful when the number of
players makes enumeration impractical but arbitrary hybrid coalitions can still
be evaluated.

\subsection{Comparison of attribution methods}

Table~\ref{tab:method-comparison} compares the allocation rule, implementation
requirements, dominant computational work, and principal tradeoff of each
method. The conventional walk is included as the operational benchmark. The
cost expressions assume that the two endpoint forecasts are already available.

\begin{table*}[htbp]
  \centering
  \caption{Comparison of methods for attributing the gap between two forecast
  runs.}
  \label{tab:method-comparison}
  \scriptsize
  \setlength{\tabcolsep}{3.5pt}
  \renewcommand{\arraystretch}{1.18}
  \resizebox{\textwidth}{!}{%
  \begin{tabular}{@{}p{2.15cm}p{3.15cm}p{3.55cm}p{2.55cm}p{5.0cm}@{}}
    \toprule
    Method & Allocation and order property & Main requirements & Dominant work
    & Best use and principal limitation \\
    \midrule
    Conventional walk
    & Exact decomposition along one selected order; generally order dependent
    & Black-box evaluations of a sequence of executable hybrid runs
    & $(k-1)T_F$
    & Simplest and easiest to communicate, but interaction effects are assigned
      according to an arbitrary order. \\

    Exact Shapley value
    & Unrestricted Shapley allocation; exact, efficient, and order independent
    & Every hybrid coalition must be executable and economically meaningful
    & $(2^k-2)T_F$
    & Preferred benchmark for a small number of input blocks; exponential cost
      limits scalability. \\

    Hierarchical/Nested Shapley
    & Owen allocation over hierarchy-respecting orders; exact and efficient,
      but dependent on the prespecified grouping
    & Economically justified hierarchy and executable hierarchy-respecting
      hybrids
    & $(Q_{\mathcal G}-2)T_F$
    & Useful when inputs have genuine group structure or dependencies; it is
      generally not the unrestricted Shapley value. \\

    Integrated Gradients
    & Straight-path Aumann--Shapley allocation; complete under exact integration
      but dependent on the selected path
    & Differentiable system and meaningful continuous interpolation between runs
    & $M T_{\nabla F}$
    & Computationally attractive for continuous inputs; unsuitable for many
      discrete changes, code replacements, and bucketed transformations. \\

    Gradient SHAP
    & Expected-gradient allocation over sampled baselines and path points;
      baseline-distribution dependent
    & Differentiable system, meaningful baselines, and valid interpolation
    & $R T_{\nabla F}$
    & Incorporates reference uncertainty; explains the specific two-run gap
      only when the baseline distribution is concentrated on the first run. \\

    Permutation SHAP
    & Monte Carlo estimate of the unrestricted Shapley value; efficiency holds
      for the sampled-walk average
    & Executable linked hybrid runs for each sampled order
    & $B(k-1)T_F$
    & Flexible, transparent, and auditable; strong interactions may require many
      walks, so sampling uncertainty must be reported. \\

    Kernel SHAP
    & Shapley-kernel weighted-regression estimate; efficiency imposed as a
      constraint
    & Executable arbitrary hybrid coalitions and a well-balanced regression
      design
    & $M T_F$
    & Supports independent, parallel coalition evaluations and a flexible
      budget; inadequate sampling can yield an unstable or design-sensitive fit. \\
    \bottomrule
  \end{tabular}
  }
\end{table*}

The cost expressions in Table~\ref{tab:method-comparison} separate expensive
forecast or gradient evaluations from lower-cost aggregation. Exact Shapley
has exponential cost in the number of players, whereas Permutation SHAP,
Kernel SHAP, Integrated Gradients, and Gradient SHAP permit an adjustable
evaluation budget. Hierarchical Shapley can reduce the coalition count when a
valid grouping is available, but the reduction reflects a restricted
allocation rule rather than a faster calculation of the unrestricted Shapley
value. Caching and parallel execution can reduce repeated work and elapsed
time for all coalition-based methods.

No method dominates across all CCAR and CECL applications. The comparison
depends on whether hybrid configurations are valid, whether differentiable
paths exist, how many input blocks are material, and whether the governance
objective requires the unrestricted Shapley value. The reported cost formulas
are leading-order approximations and exclude data preparation, model
validation, failed-run remediation, storage, and review. Method selection
should therefore consider operational feasibility and attribution semantics in
addition to the nominal number of forecast evaluations.

\section{Results}

\section{Discussion}

\section{Conclusion}

\section*{Declaration of competing interest}

The authors declare that they have no known competing financial interests or
personal relationships that could have appeared to influence the work reported
in this paper.

\section*{Data availability}

Data availability information will be added here.

\bibliographystyle{elsarticle-num}
\bibliography{references}

\end{document}